# Trigonometric Rosen-Morse Potential as the Quark-Antiquark Interaction Potential for Meson Properties in the Non-Relativistic Quark Model Using EAIM

M. Abu-Shady[1], Sh. Y. Ezz-Alarab[1]

Department of Mathematics and Computer Sciences, Faculty of Science, Menoufia University, Egypt[1]

**Abstract**

Trigonometric Rosen-Morse potential is suggested as the quark-antiquark interaction potential for studying the thermodynamic properties and the masses of heavy and heavy-light mesons. For this purpose, N-radial Schrödinger equation is analytically solved using an exact analytical iteration method (EAIM). The eigenvalues of energy and corresponding wave functions are obtained in the N-dimensional space. The present results are applied for calculating the mass of heavy mesons such as charmonium $c\bar{c}$, bottomonium $b\bar{b}$, $b\bar{c}$, and $c\bar{s}$ mesons and thermodynamic properties such as the mean-internal energy, the specific heat, the free energy, and the entropy. The effect of dimensional number is studied on the meson masses. A comparison is studied with other works and experimental data. The present potential provides satisfying results in comparison with other works and experimental data.

## 2-Introduction

The quantum chromodynamics (QCD) is an acceptable theory for describing the strong interactions of quarks that are the fundamental constituents of hadrons. The non-abelian gauge theory with the gluons is as gauge bosons and quark interactions run from one to many exchanges gluon self-gluon interactions. The theory of QCD has quark confinement feature, where quarks still very trapped but behave as free particles at high energies and momenta [1].

The Schrödinger equation (SE) plays an important role for describing many phenomena as in high energy physics. Thus, the solutions of the SE are important for calculating mass of quarkonia and thermodynamic properties. To obtain the exact and approximate solutions of SE, the various methods have been used for specific potentials such as Nikiforov–Uvarov (NU) method [2-4]. In addition, the analytical iteration method (AIM) reproduces the exact solutions to many differential equations which are important in the applications to many problems in physics, such as the equations of Hermite, Laguerre, Legendre, and Bessel. The AIM also gives a complete exact solution of Schrödinger equation for Pösch-Teller potential, the harmonic oscillator potential, the complex cubic, quartic [5], and sextic anharmonic oscillator potentials [6-8].

Recently, trigonometric Rosen-Morse potential (TRM) plays a role as an effective potential for QCD. In Refs. [9, 10], the potential is solved in one dimensions space. In Ref. [4], the authors extended the study to D-dimensional space using NU method. In Ref. [1], trigonometric quark confinement potential is considered to provide an efficient tool for quark model calculations of spectroscopic characteristics of baryons.

Therefore, the aim of this work is to extend the TRM to calculate heavy-meson properties in the free and hot media that are not considered in the previous works, in particular, spectra of heavy-meson masses and thermodynamic properties. For this purpose, the N-radial Schrödinger equation (SE) is analytically solved using the exact analytical iteration method (EAIM) for the present potential.

The paper is organized as follows: In Sec. 2, the shape of the trigonometric Rosen-Morse potential is studied. In Sec. 3, the energy eigenvalues and the corresponding wave functions are calculated in the *N*-dimensional space with TRM potential. In Sec. 4, thermodynamic properties are calculated. In Sec. 5, the results are discussed. In Sec. 6, summary and conclusion are presented.

## 2- The Trigonometric Rosen- Morse Potential (TRM)

In this section, we discuss the features of Rosen- Morse potential that takes the following form as in Refs. [1,11,12].

$$V(z) = \frac{1}{2\mu d^2}\left(-2b \cot[z] + \frac{a(a+1)}{(Sin[z])^2}\right), \qquad (1)$$

where, a=1, 2, 3,… and $z = r/d$. $\mu$, b, d are parameters will be determined later. It is quite instructive to expand the potential in a Taylor series for small z. The potential in Eq. (1), takes the following form

$$V(r) = -\frac{A}{r} + B\, r + \frac{C}{r^2} + D\, r^2, \qquad (2)$$

where, $A = \dfrac{b}{\mu d}$, $B = \dfrac{b}{3\mu d^3}$, $C = \dfrac{a(a+1)}{2\mu}$, $D = \dfrac{a(a+1)}{30\mu d^4}$.

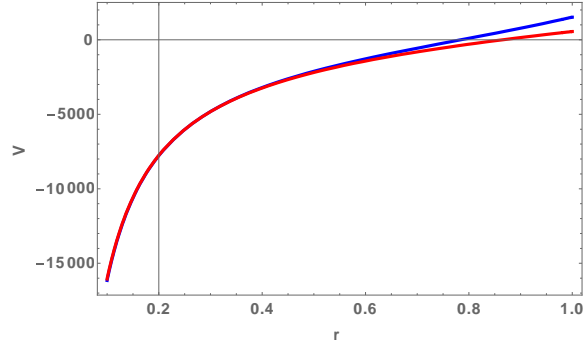

**Fig. (1).** The trigonometric Rosen-Morse potential is plotted as a function of distance r for the exact and the approximate potentials for $d = 0.5\ fm,\ b = 500,\ and\ a = 1$.

**In Fig. (1),** we note that TRM potential has two features the Coulomb potential and the confinement potential, the Coulomb potential describes the short distances and the confinement potential describes the long distances. The approximate potential closes with the exact potential up to 0.8 fm. In Ref. [1], the authors show that TRM has features of QCD for short and long distances. Thus, the approximate potential is a good potential for the description of quark-antiquark interaction potential

### 3-Exact Solution of Schrödinger equation with Trigonometric Rosen- Morse potential (TRM)

The Schrödinger equation for two particles interacting via symmetric potential in the *N*-dimensional space takes the form as in Ref. [13].

$$[\frac{d^2}{dr^2} + \frac{N-1}{r}\frac{d}{dr} - \frac{L(L+N-2)}{r^2} + 2\mu\ (E - V(r))\ ]\ \Psi\ (r) = 0, \qquad (3)$$

where L, *N,* and μ are the angular momentum quantum number, the dimensionality number, and the reduced mass for the quarkonium particle, respectively. Setting wave function $\Psi(r) = r^{\frac{1-N}{2}} R(r)$, then Eq. (3) takes the following form

$$[\frac{d^2}{dr^2} + 2\mu\ (E_{nl} - V(r)) - \frac{(L+\frac{N-2}{2})^2 - \frac{1}{4}}{2\mu\ r^2}\ ]R(r)=0, \qquad (4)$$

the trigonometric Rosen- Morse potential V(r) is given as follows as in Ref. [1]

$$V(r) = -\frac{A}{r} + B\ r + \frac{C}{r^2} + D\ r^2, \qquad (5)$$

where $A = \dfrac{b}{\mu d}$, $B = \dfrac{b}{3\mu d^3}$, $C = \dfrac{a(a+1)}{2\mu}$, and $D = \dfrac{a(a+1)}{30\mu d^4}$. (6)

By substituting Eq. (5) into Eq. (4), we obtain the following equation.

$$[\frac{d^2}{dr^2} + 2\mu(E_{nl} + \frac{A}{r} - Br - \frac{C}{r^2} - Dr^2) - \frac{(L+\frac{N-2}{2})^2 - \frac{1}{4}}{2\mu r^2}]R(r)=0, \qquad (7)$$

$$[\frac{d^2}{dr^2} + (2\mu E_{nl} + \frac{2\mu A}{r} - 2\mu Br - \frac{2\mu C}{r^2} - 2\mu Dr^2) - \frac{(L+\frac{N-2}{2})^2 - \frac{1}{4}}{r^2}]R(r)=0, \qquad (8)$$

Eq. (8) is reduced to the following form

$$[\frac{d^2}{dr^2} + (\varepsilon_{nl} + \frac{C_1}{r} - b_1 r - \frac{d_1}{r^2} + a_1 r^2)]R(r)=0, \qquad (9)$$

where,

$$\varepsilon_{nl} = 2\mu E_{nl}, \quad C_1 = 2\mu A, \quad b_1 = 2\mu B,$$

$$d_1 = 2\mu C + \left(L + \frac{N-2}{2}\right)^2 - \frac{1}{4}, \quad a_1 = -2\mu D. \qquad (10)$$

The analytical exact iteration method (AEIM) requires making the following ansatz [15] as follows

$$R_{nl}(r) = f_n(r) \exp[g_l(r)], \qquad (11)$$

where,

$$f_{n(r)} = \begin{cases} 1, & n = 0, \\ \prod_{i=1}^{n}(r - \alpha_i^{(n)}) & n = 1,2,\dots, \end{cases}$$

and,

$$g_l(r) = -\frac{1}{2}\alpha r^2 - \beta r + \delta \ln r, \qquad \alpha > 0, \beta > 0 \qquad (12)$$

It is clear that $f_n(r)$ are equivalent to the Leguere polynomial. Substituting Eq. (11) into Eq. (9), we obtain

$$R''(r) = [g_l''(r) + g_l'^2(r) + \frac{f''(r) + 2f'(r)g'(r)}{f(r)}]R_{nl}(r). \qquad (13)$$

By comparing Eq. (13) and Eq. (9), we get

$$-a_1 r^2 + b_1 r - \frac{C_1}{r} + \frac{d_1}{r^2} - \varepsilon_{nl} = g_l''(r) + g_l'^2(r)$$
$$+ \frac{f''(r) + 2f'(r) g'(r)}{f(r)}. \tag{14}$$

At n= 0, Eq. (14) takes the following form

$$-a_1 r^2 + b_1 r - \frac{C_1}{r} + \frac{d_1}{r^2} - \varepsilon_{0l} = \alpha^2 r^2 + 2 \alpha \beta r - \alpha [1+2(\delta+0)] + \beta^2$$
$$- \frac{2\beta (\delta+0)}{r} + \frac{\delta (\delta-1)}{r^2}. \tag{15}$$

Comparing the corresponding powers of r on both sides of Eq. (15), we get

$$-a_1 = \alpha^2, \; b_1 = 2 \alpha \beta \;\rightarrow\; , \beta = \frac{b_1}{2\sqrt{a_1}}, \; C_1 = 2 \beta(\delta+0), \; d_1 = \delta (\delta - 1),$$

$$\varepsilon_{0l} = \alpha [1+2(\delta+0)] - \beta^2, \tag{16}$$

then, by comparing Eq. (16) with Eq. (10), thus we get

$$\delta = \frac{1}{2}(1 \pm l'), \tag{17}$$

where, $\; l' = \sqrt{8 \mu C + 4\left(L + \frac{N-2}{2}\right)^2}, \tag{18}$

a positive sign is taken of $\delta$ that solutions of the radial wave function at the boundaries well behaved.

Now, to get energy eigenvalue equation.

$$\varepsilon_{nl} = 2 \mu E_{nl} \;\rightarrow\; \varepsilon_{0l} = 2 \mu E_{0l} \;\rightarrow\; E_{0l} = \frac{1}{2\mu} \varepsilon_{0l}, \tag{19}$$

thus, from Eq. (19), we get the energy eigenvalue equation

$$E_{0l} = \frac{1}{2\mu}[\sqrt{a_1} + \sqrt{a_1}(1+l') - \frac{b_1^2}{4 a_1}], \tag{20}$$

but $a_1 = 2 \mu D, \; b_1 = 2 \mu B, \; C_1 = 2 \mu A,$

$C_1 = 2 \beta (\delta+0), \; \delta = \frac{1}{2}(1+l') \;\rightarrow\; C_1 = \beta (1+l'),$

and $b_1 = 2 \mu B, \; \alpha = \sqrt{a_1} \;\rightarrow\; \beta = \frac{b_1}{2\sqrt{a_1}}. \tag{21}$

By substituting Eq. (21) and Eq. (6) into Eq. (20), we get

$$E_{0l} = \frac{1}{2\mu}[\sqrt{\frac{a(a+1)}{15\,d^4}}\,(2+l') - \frac{5\,b^2}{3\,a(a+1)\,d^2}], \qquad (22)$$

also, we obtain ground state function

$$\Psi_{01}(r) = \frac{1}{r}\,R_{01}(r). \qquad (23)$$

By substituting α, β and δ from Eq. (21), respectively, together and the parameters given in Eq. (6) into Eq. (23), we finally obtain the following ground state wave function

$$\Psi_{0l}(r) = N_{ol}\,r^{\frac{(-1+l')}{2}}\,e^{(-\sqrt{\frac{a(a+1)}{60\,d^4}}\,r^2 - \sqrt{\frac{5\,b^2}{3\,a(a+1)\,d^2}})}. \qquad (24)$$

Secondly, the first node (n=1), using $f_1(r) = (r - \alpha_1^{(1)})$ and $g_l(r)$ from Eq. (12), Eq. (14) is written as follows

$$-a_1 r^2 + b_1 r - \frac{C_1}{r} + \frac{d_1}{r^2} - \varepsilon_{11} = -\alpha[1+2(\delta+1)] + \alpha^2 r^2 + \beta^2 + \frac{\delta(\delta-1)}{r^2} + 2\alpha\beta\,r - \frac{2\beta\delta}{r} - \frac{2[\beta+\alpha\alpha_1^{(1)}]}{r-\alpha_1^{(1)}}] + \frac{2\delta}{r(r-\alpha_1^{(1)})}. \qquad (25)$$

By multiplying both sides of Eq. (25) by $(r - \alpha_1^{(1)})$, the relationship between the potential parameters and coefficient α, β, δ, and $\alpha_1^{(1)}$ are

$$-a_1 r^2 + b_1 r - \frac{C_1}{r} + \frac{d_1}{r^2} - \varepsilon_{11} = \alpha^2 r^2 + 2\alpha\beta\,r + \beta^2 - \alpha[1+2(\delta+1)] - \frac{2\beta(\delta+1)+\alpha\alpha_1^{(1)}]}{r} + \frac{\delta(\delta-1)}{r^2}. \qquad (26)$$

Therefore, for the relations between the potential parameters and the coefficient α, β, δ, and $\alpha_1^{(1)}$ are calculated, as by comparing both sides of Eq. (26), then $\alpha = \sqrt{a_1}$, $\beta = \frac{b_1}{2\sqrt{a_1}}$, $a_1 > 0$, $\delta = \frac{1}{2}(1+l')$,

$\varepsilon_{11} = \alpha[1+2(\delta+1)] - \beta^2$, $C_1 - 2\beta(\delta+1) = 2\alpha\alpha_1^{(1)}$, and

$$d_1 = -\delta(\delta-1) - C_1. \qquad (27)$$

Hence, the energy eigenvalue is

$$E_{11} = \frac{1}{2\mu}[\sqrt{\frac{a(a+1)}{15\,d^4}}\,(4+l') - \frac{5\,b^2}{3\,a(a+1)\,d^2}]. \qquad (28)$$

The corresponding wave function

$$\Psi_{11}(r) = N_{11} \ (r - \alpha_1^{(1)}) \ r^{\frac{(-1+l')}{2}} \ e^{(-\frac{1}{2}\sqrt{\frac{a(a+1)}{15\,d^4}} \ r^2 - \frac{b}{3\,d\,\sqrt{a\,(a+1)}} r)}. \tag{29}$$

Then, the analytical procedures node (n=2) with

$f_2(r) = (r - \alpha_1^{(2)}) (r - \alpha_2^{(2)})$ and $g_l(r)$ as defined in Eq. (11), we obtain at n = 2, l = 1, thus

$$-a_1 r^2 + b_1 r - \frac{C_1}{r} + \frac{d_1}{r^2} - \varepsilon_{21} = -\alpha[1 + 2(\delta+1)] + \alpha^2 r^2 + \beta^2 + \frac{\delta(\delta-1)}{r^2} + 2$$
$$\alpha\,\beta\,r - \frac{2\,\beta\,\delta}{r} - \frac{2+2\left(-\alpha r + \beta + \frac{\delta}{r}\right)(2\,r - \alpha_1^{(2)} - \alpha_2^{(2)})}{(r-\alpha_1^{(2)})(r-\alpha_2^{(2)})}, \tag{30}$$

By multiplying both sides of Eq. (30), by $(r - \alpha_1^{(2)})(r - \alpha_2^{(2)})$,

we get

$$-a_1 r^2 + b_1 r - \frac{C_1}{r} + \frac{d_1}{r^2} - \varepsilon_{21} = \alpha^2 r^2 + 2\,\alpha\,\beta\,r + \beta^2 - \alpha[1+2(\delta+2)] -$$
$$\frac{2\,\beta\,(\delta+2) + 2\alpha(\alpha_1^{(2)} + \alpha_2^{(2)})}{r} + \frac{\delta(\delta-1)}{r^2}, \tag{31}$$

then the relationship between the potential parameters coefficients

$\alpha, \beta, \delta, \alpha_1^{(2)}$, and $\alpha_2^{(2)}$ are $\alpha = \sqrt{a_1}$, $\beta = \frac{b_1}{2\sqrt{a_1}}$, $a_1 > 0$,

$\delta = \frac{1}{2}(1+l')$, $\varepsilon_{21} = \alpha[1+2(\delta+2)] - \beta^2$,

$$C_1 - 2\,\beta\,(\delta + 2) = 2\,\alpha \sum_{i=1}^{2} \alpha_i^{(2)}. \tag{32}$$

Hence, the energy eigenvalue

$$E_{21} = \frac{1}{2\,\mu} \left[ \sqrt{\frac{a(a+1)}{15\,d^4}} \,(6 + l') - \frac{5\,b^2}{3\,a(a+1)} \right]. \tag{33}$$

The corresponding wave function is

$$\Psi_{21}(r) = N_{21} \ \prod_{i=1}^{2}(r - \alpha_i^{(2)}) \ r^{\frac{(-1+l')}{2}} \ e^{\left(-\frac{1}{2}\sqrt{\frac{a(a+1)}{15\,d^4}} \ r^2 - \frac{b\sqrt{15}}{3\,d\,\sqrt{a\,(a+1)}} r\right)}. \tag{34}$$

The present method is applied for the third node (n=3) by taking

$f(r) = (r - \alpha_1^{(3)})(r - \alpha_2^{(3)})(r - \alpha_3^{(3)})$, and $g_l(r)$ are defined in Eq. (11), to obtain

$$- a_1 r^2 + b_1 r - \frac{C_1}{r} + \frac{d_1}{r^2} - \varepsilon_{31} = \alpha^2 r^2 + 2\alpha\beta r + \beta^2 - \alpha[1+2(\delta+3)] -$$
$$\frac{2[\beta(\delta+3)+\alpha\sum_{i=1}^{3}\alpha_i^{(3)}]}{r} + \frac{\delta(\delta-1)}{r^2}, \tag{35}$$

then, the relationship between the potential parameters coefficients

$\alpha, \beta, \delta, \alpha_1^{(3)}, \alpha_2^{(3)},$ and $\alpha_3^{(3)}$ are $\alpha = \sqrt{a_1}$, $\beta = \frac{b_1}{2\sqrt{a_1}}$, $a_1 > 0$,

$\delta = \frac{1}{2}(1+l')$, $\varepsilon_{31} = \alpha[1+2(\delta+3)] - \beta^2$,

$$C_1 - 2\beta(\delta+3) = 2\alpha \sum_{i=1}^{3}\alpha_i^{(3)}. \tag{36}$$

The coefficient $\alpha_1^{(3)}$, $\alpha_2^{(3)}$, and $\alpha_3^{(3)}$ are found from the constraint relation,

$$\alpha \sum_{i=1}^{3}\alpha_i^{(3)^2} + \beta \sum_{i=1}^{3}\alpha_i^{(3)} - 3(\delta+1) = 0, \tag{37}$$

the energy eigenvalue formula is

$$E_{31} = \frac{1}{2\mu}\left[\sqrt{\frac{a(a+1)}{15 d^4}}(8+l') - \frac{5b^2}{3a(a+1)d^2}\right]. \tag{38}$$

The corresponding wave function is

$$\Psi_{31}(r) = N_{31} \prod_{i=1}^{3}(r - \alpha_i^{(3)}) \; r^{\frac{(-1+l')}{2}} \; e^{\left(-\frac{1}{2}\sqrt{\frac{a(a+1)}{15 d^4}} r^2 - \frac{b\sqrt{15}}{3d\sqrt{a(a+1)}} r\right)}. \tag{39}$$

We can repeat this iteration procedure several times to write the exact energy formula for the TRM potential with any arbitrary n and l state as

$$E_{nl} = \frac{1}{2\mu}\left[\sqrt{\frac{a(a+1)}{15 d^4}}(2+2n+l') - \frac{5b^2}{3a(a+1)d^2}\right]. \tag{40}$$

The corresponding wave function for any n state is

$$\Psi_{nl}(r) = N_{nl} \prod_{i=1}^{n}(r - \alpha_i^{(n)}) \; r^{\frac{(-1+l')}{2}} \; e^{\left(-\frac{1}{2}\sqrt{\frac{a(a+1)}{15 d^4}} r^2 - \frac{b\sqrt{15}}{3d\sqrt{a(a+1)}} r\right)}. \tag{41}$$

Then, the relationship between the potential parameters coefficients

$\alpha$, $\beta$, $\delta$, $\alpha_1^{(n)}$, $\alpha_2^{(n)}$, ..., $\alpha_3^{(n)}$ are $\alpha=\sqrt{a_1}$, $\beta = \frac{b_1}{2\sqrt{a_1}}$, $a_1 > 0$,

$\delta = \frac{1}{2}(1+ l')$, $\varepsilon_{nl} = \alpha[1+ 2(\delta+3)] - \beta^2$,

$C_1 - 2\beta(\delta + n) = 2\alpha \sum_{i=1}^{n} \alpha_i^{(n)}$, n=1, 2, 3, (42)

the coefficient $\alpha_1^{(3)}$, $\alpha_2^{(3)}$, and $\alpha_3^{(3)}$ are found from the constraint relation,

$\alpha \sum_{i=1}^{3} \alpha_i^{(3)^2} + \beta \sum_{i=1}^{3} \alpha_i^{(3)} - 3(\delta + 1) = 0$. (43)

**4- Thermodynamics properties**

In this section, thermodynamic properties of the Trigonometric Rosen-Morse potential are studied, the partition function is given $Z=\sum_{n=0}^{\infty} e^{-\beta E}$, where $\beta = \frac{1}{KT}$, K is the Boltzmann constant as in Ref. [2]

**(4.1)-Partition function**

$Z(\beta)=\sum_{n=0}^{\infty} e^{-\beta E_{nl}}$, (44)

by substituting Eq. (40), we obtain

$Z(\beta) = \frac{e^{-\beta(C_1+C_2-C_3)}}{(1-e^{-\beta C_1})}$, (45)

where, $C_1 = \frac{1}{\mu}\sqrt{\frac{a(a+1)}{15d^4}}$, $C_2 = \frac{l'}{2\mu}\sqrt{\frac{a(a+1)}{15d^4}}$, $C_3 = \frac{5}{6\mu}\frac{b^2}{a(a+1)d^2}$,

$l' = \sqrt{8\mu C + 4\left(L + \frac{N-2}{2}\right)^2}$. (46)

**(4.2)- Mean energy U**

$U(\beta) = -\frac{d}{d\beta} \text{Ln}Z(\beta)$, (47)

$U(\beta)$
$= -e^{\beta(C_1+C_2-C_3)}(1 - e^{-\beta C_1})\left(-\frac{e^{-\beta C_1-\beta(C_1+C_2-C_3)}C_1}{(1-e^{-\beta C_1})^2}\right.$
$\left. + \frac{e^{-\beta(C_1+C_2-C_3)}(-C_1 - C_2 + C_3)}{1 - e^{-\beta C_1}}\right)$. (48)

### (4.3)- Specific heat C

$$C(\beta) = \frac{dU}{dT} = -K\beta^2 \frac{dU}{d\beta}, \tag{49}$$

$$
\begin{aligned}
C(\beta) &= -K\beta^2 \Big( -e^{-\beta C_1 + \beta(C_1+C_2-C_3)} C_1 \Big( -\frac{e^{-\beta C_1 - \beta(C_1+C_2-C_3)} C_1}{(1-e^{-\beta C_1})^2} \\
&+ \frac{e^{-\beta(C_1+C_2-C_3)}(-C_1-C_2+C_3)}{1-e^{-\beta C_1}} \Big) - e^{\beta(C_1+C_2-C_3)}(1-e^{-\beta C_1})(C_1+C_2 \\
&- C_3) \Big( -\frac{e^{-\beta C_1 - \beta(C_1+C_2-C_3)} C_1}{(1-e^{-\beta C_1})^2} + \frac{e^{-\beta(C_1+C_2-C_3)}(-C_1-C_2+C_3)}{1-e^{-\beta C_1}} \Big) \\
&- e^{\beta(C_1+C_2-C_3)}(1-e^{-\beta C_1}) \Big( \frac{2e^{-2\beta C_1 - \beta(C_1+C_2-C_3)} C_1^2}{(1-e^{-\beta C_1})^3} \\
&- \frac{e^{-\beta C_1 - \beta(C_1+C_2-C_3)} C_1(-2C_1-C_2+C_3)}{(1-e^{-\beta C_1})^2} \\
&- \frac{e^{-\beta C_1 - \beta(C_1+C_2-C_3)} C_1(-C_1-C_2+C_3)}{(1-e^{-\beta C_1})^2} \\
&+ \frac{e^{-\beta(C_1+C_2-C_3)}(-C_1-C_2+C_3)^2}{1-e^{-\beta C_1}} \Big) \Big), \tag{50}
\end{aligned}
$$

### (4.4)- Free energy

$$F(\beta) = -KT\,LnZ(\beta), \tag{51}$$

$$F(\beta) = -\frac{Log\big[\frac{e^{-\beta(C_1+C_2-C_3)}}{(1-e^{-\beta C_1})}\big]}{\beta} \tag{52}$$

### (4.5)- Entropy

$$S(\beta) = K \ln Z(\beta) - K\beta \frac{\partial}{\partial \beta} \ln Z(\beta), \tag{53}$$

$$S(\beta) = KLog\big[\frac{e^{-\beta(C_1+C_2-C_3)}}{1-e^{-\beta C_1}}\big] - e^{\beta(C_1+C_2-C_3)}(1-e^{-\beta C_1})K\beta\Big(-\frac{e^{-\beta C_1-\beta(C_1+C_2-C_3)}C_1}{(1-e^{-\beta C_1})^2} + \frac{e^{-\beta(C_1+C_2-C_3)}(-C_1-C_2+C_3)}{1-e^{-\beta C_1}}\Big). \tag{54}$$

## 5. Results and Discussion
### [5.1] Quarkonium masses

In this section, we calculate spectra of the heavy quarkonium system such as charmonium and bottomonium that have the quark and antiquark flavor, the mass of quarkonium is calculated in 3-dimensional space ($N = 3$). So we apply the following relation as in Refs. [14-15].

$$M = 2m + E_{nl}, \qquad (56)$$

where, m is bare quark mass for quarkonium. By using Eq. (41), we can write Eq. (56) as follows

$$M = 2m + \frac{1}{2\mu}\left[\sqrt{\frac{a(a+1)}{15\, d^4}}(2 + 2n + l') - \frac{5\, b^2}{3\, a(a+1)\, d^2}\right], \qquad (57)$$

In Table (1), we calculated mass spectra of charmonium for states from 1S to 2D, this by using Eq. (57). The free parameters of the present calculations are A, B, C and D are fitted with experimental data. In addition, quark masses are obtained from Ref. [27]. We note that calculation of masses of charmonium are in a good agreement with experimental data and are improved in comparison with Refs. [16, 17, 19, 20, 22] by calculating total error in comparison with experimental data. In Table (2), we note the spectra masses of bottomonium from states 1S to 4S using Eq. (57) are in agreement with experimental data, and the present calculations are improved in comparison with Refs. [16-17,18,20] in which the total error is reduced in comparison with these works. In Table (3), we calculate mass spectra of meson b$\bar{c}$ mesons from using $2m = m_b + m_c$ part of Eq. (57) for states from 1S to 3S. We find that the 1S state closes with experimental data, but the values of the experimental data for other states are not available. We note that the present calculations of the b$\bar{c}$ mass improved in comparison with Refs. [23-25] by calculating the total error for these works. In Ref. [16], the authors used the asymptotic iteration method to solve the SE, in which the Cornell potential is extended to the harmonic oscillator. Also, In Ref. [19], the authors used the same potential as in Ref. [16] and used NU method for solving SE. Thus, the potential that was employed in Refs. [16, 17] is particular case from the present potential at C = 0. In addition, the calculations in Refs. [16, 17] are carried out for charmonium and bottomonium only. In Refs. [22, 23, 24], the authors employed the Cornell potential. Thus, the potential in Refs. [22, 23, 24] is a particular case from the present potential at C = D = 0. In Ref. [17], the authors employed the similar potential with using Laplace transformation method. Thus, the present method gives better results in comparison with Ref. [17].

In Table (4), we calculate mass spectra of c$\bar{s}$ mesons from 1S to 1D state, by using 2m=$m_c$+$m_s$ part of Eq. (57). 1S and 1D close with experimental data. Other states are improved in comparison with power potential, screened potential, and phenomenological potential [26] by calculating the total error for each potential. Thus, we deduce that TRM gives good results for charmonium, bottomonium, b$\bar{c}$, and c$\bar{s}$ meson in comparison with experimental data and improved in comparison with recent works.

The higher dimensional space plays an important role in the particles physics. Based on superstring theory such as Ref. [4] and references therein, the number of dimensions in the universe restricted to ten spatial dimensions and one for the time dimension. If the amount of spatial dimensional increases more than ten, the universe unstable and collapse. In present work, we note that the energy eigenvalues increase with increasing the dimensionality number at N = 4 for every state as in Tables (1, 2, 3, 4). Therefore, the spectra of states of masses increase with increasing dimensionality number. This lead to the limitation of non-relativistic quark models when we apply on the heavy quark meson.

**Table (1)**. Mass spectra of charmonium (in GeV), ( $m_c$=1.209 GeV,

$\mu$ =0.6042 GeV, d = 1.6869 GeV$^{-1}$, b = 2.08981 GeV$^2$, a =2.13.

| State | Present paper | [16] | [17] | [19] | [20] | [22] | N=4 | Exp. |
|---|---|---|---|---|---|---|---|---|
| 1s | 3.148 | 3.078 | 3.096 | 3.096 | 3.078 | 3.096 | 3.360 | 3.096 |
| 1p | 3.286 | 3.415 | 3.433 | 3.433 | 3.415 | 3.255 | 3.673 | - |
| 2s | 3.536 | 4.187 | 3.686 | 3.686 | 3.581 | 3.686 | 3.698 | 3.686 |
| 1D | 3.522 | 3.752 | 3.767 | 3.770 | 3.749 | 3.504 | 3.895 | - |
| 2p | 3.674 | 4.143 | 3.910 | 4.023 | 3.917 | 3.779 | 3.827 | 3.773 |
| 3s | 3.924 | 5.297 | 3.984 | 4.0404 | 4.085 | 4.040 | 3.966 | 4.040 |

| | | | | | | | |
|---|---|---|---|---|---|---|---|
| 4s | 4.311 | 6.407 | 4.150 | 4.355 | 4.589 | 4.269 | 3.986 | 4.263 |
| 2D | 3.909 | - | - | 3.096 | 3.078 | - | 4.170 | 4.159 |
| Total Error | 0.18378 | 0.59996 | 0.07668 | 0.34343 | 0.41999 | 0.42 | | |

**Table (2).** Mass spectra of bottomonium (in GeV), ( $m_b$= 4.823 GeV, $\mu$ =2.4115 GeV, $d = 1.101888$ GeV$^{-1}$, b = 3.55007 GeV$^2$, $a = 2$.

| State | Present paper | [16] | [20] | [17] | [18] | N=4 | Exp. |
|---|---|---|---|---|---|---|---|
| 1s | 9.524 | 9.510 | 9.510 | 9.460 | 9.460 | 9.610 | 9.460 |
| 1p | 9.885 | 9.862 | 9.862 | 9.840 | 9.811 | 10.022 | - |
| 2s | 10.02 | 10.627 | 10.038 | 10.023 | 10.023 | 10.072 | 10.023 |
| 1D | 10.02 | 10.214 | 10.214 | 10.140 | 10.161 | 10.205 | - |
| 2p | 10.01 | 10.944 | 10.390 | 10.160 | 10.374 | 10.269 | - |
| 3s | 10.236 | 11.726 | 10.566 | 10.280 | 10.355 | 10.306 | 10.355 |
| 4s | 10.452 | 12.834 | 11.094 | 10.420 | 10.655 | 10.344 | 10.580 |
| Total Error | 0.02953 | 0.941149 | 0.075741 | 0.0323 | 0.00701 | | |

**Table(3).** Mass spectra of $b\bar{c}$ meson (in GeV) ($m_b$=4.823 GeV, $m_c$=1.209 GeV, $\mu = 0.9666788$ GeV, $d = 0.94984$ GeV$^{-1}$, $a =1.5$, $b = 2.405923$ GeV$^2$ .

| State | Present paper | [23] | [24] | [25] | N=4 | Exp. |
|---|---|---|---|---|---|---|
| 1s | 6.277 | 6.349 | 6.264 | 6.270 | 6.355 | 6.277 |
| 1p | 6.535 | 6.715 | 6.700 | 6.699 | 6.883 | - |
| 2s | 6.850 | 6.821 | 6.856 | 6.835 | 6.878 | - |
| 2p | 7.108 | 7.102 | 7.108 | 7.091 | 7.161 | - |
| 3s | 7.423 | 7.175 | 7.244 | 7.193 | 8.035 | - |
| Total Error | 0 | 0.0114704 | 0.002071 | 0.001115 | | |

**In Table**(4). Mass spectra of $c\bar{s}$ meson in (GeV) (($m_c$=1.029, $m_s$=0.419) GeV, $\mu = 0.31116$ GeV, $d = 2.39835$ GeV$^{-1}$, $a =1.5$ $b = 2.39835$ GeV$^2$,

| State | Present paper | power | Screened | Phenomenological | N=4 | Exp. |
|---|---|---|---|---|---|---|
| 1s | 1.968 | 1.9724 | 1.9685 | 1.968 | 2.3001 | 1968.3[30] |
| 1p | 2.313 | 2.540 | 2.7485 | 2.566 | 2.742 | - |
| 2s | 2.735 | 2.6506 | 2.8385 | 2.815 | 2.797 | 2.709[36] |
| 3S | 3.501 | 2.9691 | 3.2537 | 3.280 | 2.967 | - |
| 1D | 3.859 | - | - | - | 3.934 | 2.859 [36] |
| Total Error | 0.0096 | 0.0238 | 0.04137 | 0.03913 | | |

## [5.2]- Thermodynamics properties

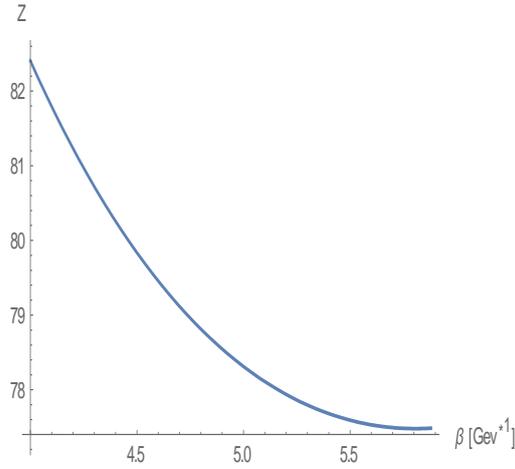

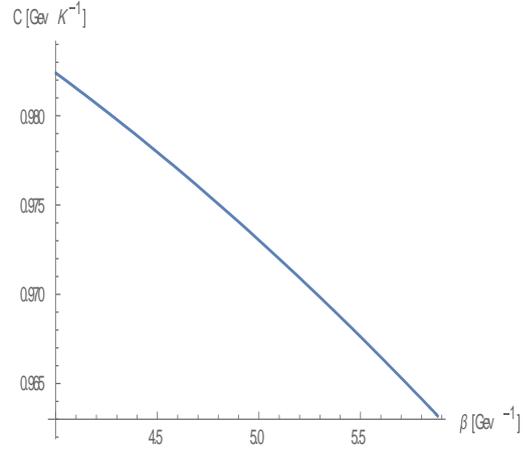

**Fig. 2:** Partition function is plotted as a function of $\beta$.    **Figure 3:** Specific heat C is plotted as a function $\beta$

In Fig. (2), we note that partition function (Z) decreases with increasing $\beta$. The range of $\beta$ = 4.0 to 5.88 Mev$^{-1}$ corresponding to T = 0.25 to 0.170 GeV which represents the range of temperature above the critical temperature. In Refs. [26, 27], the authors studied the thermodynamic properties for diatomic molecules in the relativistic models using NU method. They found that the partition function decreases with increasing $\beta$. In addition, In Ref. [28], the authors employed oscillator plus inverse square potential in the SE and found the partition function decreases with increasing $\beta$. Therefore, we found the present behavior is a qualitative agreement with these works. In Fig. (3), we note that the specific heat (C) decreases with increasing of $\beta$. We found a qualitative agreement with these works [26, 27, 28].

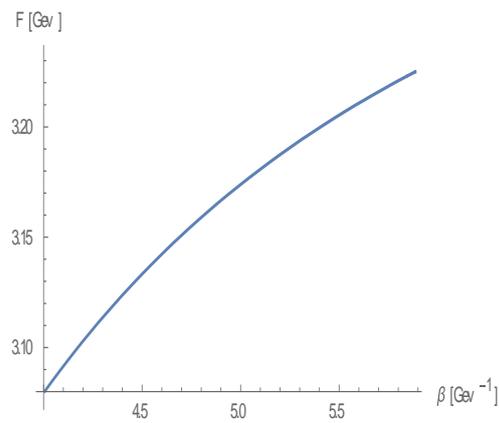

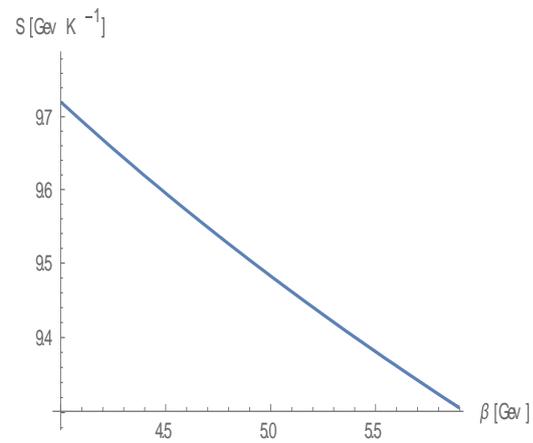

**Figure 4:** Free energy F is plotted as a function $\beta$    **Figure 5:** Entropy S is plotted as a function $\beta$

In Fig. (4), we note that the free energy (F) increases with increasing $\beta$. In Ref. [5], the free energy increases with decreasing of $\beta$. In Refs. [30-35], the free energy increases with decreasing temperature. Therefore, the present result is an agreement with Refs. [30-35]. In Fig. (5), we note that the entropy (S) decreases with increasing β. This finding is an agreement with Refs. [26-31], in which the entropy increases with increasing temperature for the diatomic molecules

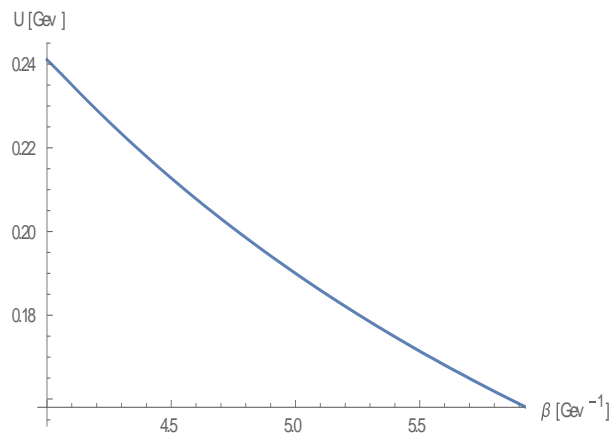

**Fig .6:** Internal energy U is plotted as a function β

**In Fig. (6),** we note that internal mean energy (U) decreases with increasing of β. In Ref. [5], the authors found that U increases with decreasing of β. This observation is noted in Ref. [30] for the diatomic molecules HCl and $H_2$. In addition, In Refs. [30, 31], U is plotted as a function of the dimensionless temperature and the authors found that U increases with increasing of temperature. Thus, the present work has the same conclusion for the charm medium.

**6- Summary and Conclusion.**

In this work, the trigonometric Rosen- Morse potential is suggested as an effective potential for quark-antiquark interaction, in which the potential satisfied the features of QCD. By using the analytical exact iteration method (AEIM), the eigenvalues of energy and corresponding wave functions are obtained in the N-dimensional radial Schrodinger equation. The present results are applied for calculating the mass of heavy mesons such as charmonium c c̄, bottomonium b b̄, b c̄, and c s̄ mesons and thermodynamic properties such as the internal energy, the specific heat, the free energy, and the entropy.

I- At N=3, the heavy meson spectra masses are calculated. We obtained the total errors as 0.18378 for charmonium, 0.02953 for bottomonium, 0 for the b $\bar{c}$ meson, and 0.0096 for the c $\bar{s}$ meson in comparison with experimental data. For N > 3, we found that the binding energy increases with increasing dimensional number space which leads to the limitation of non-relativistic quark models.

II- At N = 3, thermodynamic properties are calculated such as the internal energy, the free energy, the specific heat, and the entropy. We found thermodynamic properties for the charm quark plasma are a qualitative agreement with Refs. [30, 35]. In these works, thermodynamic properties the light quark, the strange quark, and natural particles are studied. We conclude that the trigonometric Rosen- Morse potential gives good results in comparison with other recent works and the present results are good agreement with experimental data. In addition, the TRM potential provides satisfied results for thermodynamic properties for charm medium.

# 7- References

[1] C. B. Compean and M. Kirch, Eur. Phys. J. A **33**, 1 (2007).

[2] A. N. Ikot, B. C. Lutfuoglu, M. I. Ngwueke, M. E. Udoh, S. Zare, and H. Hassan, Eur. Phys. J. Plus, 131, 419 (2016).

[3] M. Abu-Shady, T. A. Abdel-Karim, Sh. Y. Ezz-Alarab, Journal of the Egyptian Mathematical Society (Accepted) arXiv:1901.00470.

[4] U. A. Deta, Suparmi, Cari, A. S. Husein, H. Yulian, Khaled, I. K. A. Luqman, and Supriyanto, 4$^{th}$. Inter. Conf. Adv. Nucl. Sci. Eng.,**1615**,121-127(2014).

[5] H. Ciftci, R. L. Hall, and N. Saad, J. Phys. A **38**, 1147 (2005).

[6] F. M. Fern´andez J. Phys. A **37**, 6173, (2004).

[7] T. Barak, K. Abod, O.M. Al-Doss, Czecho. J. Phys. **56** , 6,(2006).

[8] A. Arda , C. Tezcan, and R. Sever, Few. Body. Sys., **57**, 101, (2016).

[9] C. B. C. Jasso and M. Kirchbach, AIP Conf. Proc. **857**, 275-278 (2006).

[10] C. B. C. Jasso and M. Kirchbach, J. Phys. A: Math. Gen. **39**, 547 (2006).

[11] C. V. Sukumar, J. Phys. A:Math. Gen. 18, 2917 (1998); AIP proceedings 744, eds. R. Bijker et al, Supersymmetries in physics and applications, 167 (New York, 2005).

[12] F. Cooper, A. Khare, and U. P. Sukhat, Super. Symmt. Quant. Mechan. (World Scientific, Singapore), (2001).

[13] M. Abu-Shady, T. A. Abdel-Karim, E. M. Khokha, Advances in High Energy Physics, 2018, 7356843 (2018).

[14] M. Abu-Shady, E. M. Khokha, Advances in High Energy Physics, 2018, 7032041 (2018).

[15] M. Abu-Shady, T. A. Abdel-Karim , E. M. Khokha, and SF. J. Quan. Phys. 2, 1000017, ( 2018).

[16] R. Kumar and F. Chand, Commun. Theor. Phys. **59**, 528 (2013).

[17] A. Al-Jamel and H. Widyan, Appl. Phys. Rese. **4**, 94 (2013).

[18] Z. Ghalenovi, A. A. Rajabi, S. Qin and H. Rischke, hep-ph/14034582 (2014).

[19] N. V. Masksimenko and S. M. Kuchin, Russ. Phy. J. **54**, 57 (2011).

[20] R. Kumar and F. Chand, Phys. Scr. **85**, 055008 (2012).

[21] Y. Li, P. Maris, and J. P. Vary. Phys. Review D: Particles, Fields, Gravit. Cosmology. 96, 1, (2017).

[22] S. M. Kuchin and N. V. Maksimenko, Univ. J. Phys. Appl. **7**, 295 (2013).

[23]A. Kumar Ray and P. C. Vinodkumar, Pramana J. Phys. **66**, 958 (2006).

[24] E. J. Eichten and C. Quigg, Phys. Rev. D **49**, 5845 (1994).

[25] D. Ebert, R. N. Faustov, and V. O. Galkin, Phys. Rev. D **67**, 014027 (2003).

[26] D. Ebert , R. N. Faustov, and V. O. Galkin , Phys. Rev. D **67**, 014027 (2003).

[27] A. Al-Jamel and H. Widyan, Appl. Phys. Rese. **4**, 94 (2013).

[28] T. Das, EJTP 13, 207 (2016).

[29] S. Roy, and D. K. Choudhury, Cana. J. Phys. **94**, 1282 (2016).

[30] W. A. Yah and K.J. Oyew, J. Asso. Arab. Univ. Bas. App. Sci. **21**, 53 (2016).

[31] M. C. Onyeaj, A. N. Ikot, C. A. Onate, O. Ebomw, M. E. Udoh, and J. O. A. Idiod, Eur. Phys. J. Plus .132, 302(2017).

[32] Shi-Hal, M. Loz-cass, J. F. Jim-Ngm, and A. L. River, Inter. J. Quant. Chem, 107, 366–371 (2007) .

[33] H. Hassan and M. Hosseinp, Eur. Phys. J. C 76, 553 (2016).

[34] M. H. Pach, R. V. Maluf, C. A. S. Almeid and R. R. Land, arXiv:1406.5114v2, 11, (2014).

[35] E. Meg, E. Ruiz and L. L. Salced, arXiv:1603.04642v3, 18, (2016).